# Generative Adversarial Learning for Trusted and Secure Clustering in Industrial Wireless Sensor Networks

Liu Yang, Simon X. Yang, *Senior Member*, *IEEE*, Yun Li, Yinzhi Lu, and Tan Guo, *Member*, *IEEE*

*Abstract*—Traditional machine learning techniques have been widely used to establish the trust management systems. However, the scale of training dataset can significantly affect the security performances of the systems, while it is a great challenge to detect malicious nodes due to the absence of labeled data regarding novel attacks. To address this issue, this paper presents a generative adversarial network (GAN) based trust management mechanism for Industrial Wireless Sensor Networks (IWSNs). First, type-2 fuzzy logic is adopted to evaluate the reputation of sensor nodes while alleviating the uncertainty problem. Then, trust vectors are collected to train a GAN-based codec structure, which is used for further malicious node detection. Moreover, to avoid normal nodes being isolated from the network permanently due to error detections, a GAN-based trust redemption model is constructed to enhance the resilience of trust management. Based on the latest detection results, a trust model update method is developed to adapt to the dynamic industrial environment. The proposed trust management mechanism is finally applied to secure clustering for reliable and real-time data transmission, and simulation results show that it achieves a high detection rate up to 96%, as well as a low false positive rate below 8%.

*Index Terms*—Industrial Wireless Sensor Networks (IWSNs), generative adversarial network (GAN), trust, security, clustering.

## I. INTRODUCTION

INDUSTRIAL Wireless Sensor Network (IWSN) is a very promising paradigm that incorporates sensor networks with industrial control systems for promoting industrial intelligence, improving production efficiency, and reducing manufacturing costs [1], [2]. The advantages provided by IWSN, such as low installation cost, scalable nature, and interoperability, have appeared to be a good reason to persuade various industrial systems to its adoption, especially in low data rate monitoring field [3]-[5]. A typical IWSN contains a huge number of sensor nodes that can collect critical parameters of industrial devices. The sensor data is ultimately delivered to a sink node and then analyzed to invoke the business process, with the aim of optimizing the automation in industrial production [6].

Due to the dynamic characteristic and application in valuable industry, IWSNs face significant challenges, such as unstable links, node faults, and security threats. These factors severely affect the reliable and real-time data delivery, which is a crucial performance requirement for the majority of industrial fields [7], [8]. The adversary tries the best to disrupt normal industrial production by destroying the accuracy, integrity, and timeliness of sensor data. Moreover, occasional node faults or link failures inevitably cause the data dropping or delaying, which is hard to be distinguished from the one due to malicious attacks. Hence, for reliable and real-time communication in industrial field, IWSNs are required to prevent against malicious attacks and keep a good tolerance to node faults and link failures.

### A. Motivation

Trust management is considered as a good method to secure IWSNs [6], [9], [10]. But how to improve the resilience of trust management has been rarely addressed before. In addition, due to absence of trust information for novel attacks and resource limitation of sensor nodes, how to establish the trust model with limited information is another research problem to be solved.

Machine learning is a powerful technique to improve system performance by using existing knowledge structure. Recently, it has been used to establish the trust management system for sensor networks and proved better adaptability to unknown or dynamic environment [11]-[13]. However, traditional machine learning algorithms are not very suitable for trust management in IWSNs, since the training dataset scale significantly affects security performance. Generative adversarial network (GAN) [14]-[16] is a typical deep learning method. It comprises a generator and a discriminator that both are trained by playing an adversarial *min-max* game. The main goal of GAN is learning potential distribution of the samples and then product data according to that distribution. Therefore, GAN has the potential to identify patterns and make proper decisions even if the training dataset scale is small, and it can also be used to predict future attacks so that suitable resilient mechanisms can be considered.

Based on the above views, the application of GAN in trust management will be explored in this paper, for the purpose of achieving reliable and real-time communication in IWSNs.

This work was supported in part by the National Natural Science Foundation of China under Grant 61801072, and in part by the Science and Technology Research Program of Chongqing Municipal Education Commission under Grant KJQN202000641. (*Corresponding authors: Simon X. Yang, Yun Li*).

Liu Yang, Yinzhi Lu, and Tan Guo are with the School of Communication and Information Engineering, Chongqing University of Posts and Telecommunications, Chongqing 400065, China (e-mail: yangliu@cqupt.edu.cn; luyinzhi@yznu.edu.cn; guot@cqupt.edu.cn).

Yun Li is with the School of Software Engineering, Chongqing University of Posts and Telecommunications, Chongqing 400065, China (e-mail: liyun@cqupt.edu.cn).

Simon X. Yang is with the Advanced Robotics and Intelligent Systems Laboratory, School of Engineering, University of Guelph, Guelph, ON N1G2W1, Canada (e-mail: syang@uoguelph.ca).

*B. Contribution*

Our main contributions are listed as follows:

1) To alleviate trust uncertainty problem, an interval type-2 fuzzy logic system is designed for fuzzy trust evaluation.

2) For trust decision-making, we present a GAN-based trust classification model, where a GAN pair is trained to construct a codec structure while conditional information is introduced to guide the model training.

3) To enhance the resilience of trust management, GANs are used to construct a trust redemption model. A codec module is used to learn sample features from context and predict future attacks, and then the false positive nodes may be trusted by others before their trust levels have been improved.

4) To adapt to the dynamic industrial environment, a trust model update method is proposed to realize adaptive retraining according to the latest trust decision-making results.

The organization of this paper is given here: First, some related works are reviewed in Section II. Our trust management method and its application in clustering are then detailed in Sections III and IV, respectively. Next, theoretical analyses are given in Section V, and experiments are conducted in Section VI. We finally conclude this work in Section VII.

## II. RELATED WORKS

Previous works on trust and security are mainly focused on three aspects, which are trust evaluation, trust classification, and anomaly detection. A brief review will be given below.

*A. Trust evaluation with fuzzy techniques*

Trust management has been widely adopted in Internet of Things (IoT) driven sensor networks. Usually, trust uncertainty problem exists since the rules to be performed and evidences to be supported are fuzzy [17]. Fuzzy logic and cloud model have been adopted to solve this problem.

Almogren *et al.* presented a trust method with fuzzy logic to prevent Sybil attacks in Internet of Medical Things (IoMT) [18]. Before providing the particular service, fuzzy inferring was performed by the server to estimate the reputation of IoMT nodes. Trust attributes such as the integrity, compatibility, and receptivity of a node towards others were used as the trust evidences. Then, malicious nodes were identified according to the predefined trust threshold. Kousar *et al.* presented a trust-aware security method with type-2 fuzzy logic for mobile sensor networks [19]. To compute the direct trust value, a type-2 fuzzy logic system was introduced, where fuzzy sets were used to assess the input and output while the trust attributes that include residual energy, distance to sink node, concentration, and moving speed were considered for inferring. The final trust result was acquired by combining the direct trust results and the recommended ones from the common neighbors. Zhang *et al.* proposed a trust evaluation method for clustered sensor networks using cloud model (TECC) [20]. First, factor trust clouds regarding communication, message, and energy were established. Then, immediate trust cloud was built by combining the weighted factor trust clouds. Next, the final trust cloud was determined through integrating the immediate and recommended clouds, and the corresponding trust grade was obtained via trust decision-making.

The presented methods in [18]-[20] can effectively alleviate the trust uncertainty since fuzzy techniques have been used for trust models construction. Mamdani type-2 fuzzy logic system and cloud model, which have been adopted in [19] and [20] respectively, can provide smoother control surface than the type-1 fuzzy logic system used in [18]. However, trust decision has to be made based on the pre-defined trust threshold for these methods, so that the adaptability to dynamic environment needs to be improved. To solve this problem, we will further adopt type-2 fuzzy logic to build the trust evaluation model, and a GAN-based trust management framework will be presented for dynamic trust decision-making.

*B. Trust classification with machine learning*

Machine learning is a powerful data exploration method for 'trustworthy' and 'untrusty' patterns learning [21]. Then it can be used to predict future novel attacks with existing examples.

Han *et al.* proposed a synergetic trust model using support vector machine (SVM) for underwater sensor networks [9]. All sensor nodes were first grouped into clusters. Trust evidences were collected by cluster members and then delivered to the cluster heads. Next, K-means algorithm was adopted to label the evidences, which were further used to train a SVM classifier. The trained classifier was finally fed back to member nodes to help for detecting the malicious. Jiang *et al.* gave a dynamic trust evaluation and update scheme using C4.5 decision tree (TEUC) for underwater sensor networks [11]. The collected evidences for trust evaluation were first normalized. Then, the evidences were converted into fuzzy sets and used to train a C4.5 decision tree. Moreover, the penalty and reward factors were introduced for trust model update according to a sliding time window. To assure the security in IoT applications, a trust computational model based on machine learning was proposed by Jayasinghe *et al.* [12]. Features of trust attributes including the experience, knowledge, and reputation were first extracted. After that, principal component analysis (PCA) and K-Means were used to perform dimension reducing and label appending respectively for trust attributes. A SVM-based trust prediction model was finally trained using the labelled data. To secure IoT systems and prevent against conditional packet manipulation attacks, a detection method with trust evaluation was proposed by Liu *et al.* [22]. A regression model was firstly established to compute trust values of IoT devices based on the relevant reputation of routing paths. Next, clustering was performed to divide trust values into three groups with different trust levels. Finally, trust decision was made based on the group to which the trust value of an IoT device belongs.

Machine learning algorithms like SVM, C4.5 decision tree, PCA, K-Means, and regression model have been successfully used to establish the trust classification models in the above methods. However, their performances are limited by the scale of training dataset, which is hard or expensive to achieve in industrial fields. The tolerance to occasional node faults and link failures has not been considered as well. To address this issue, we will use GANs to construct our trust classification model, which can alleviate the dependence on training dataset scale. GAN-based trust redemption will also be studied, for the purpose of improving the resilience of trust management.

*C. Anomaly detection with GANs*

Due to the absence of labeled data regarding novel attacks, it is with great challenges for intrusion or anomaly detection via traditional machine learning techniques. In this context, GAN is a promising unsupervised scheme to model the system. A discriminator is trained to determine whether a sample is real or fake, while an adversarial generator is trained to deceive the discriminator by synthesizing ever more realistic samples [23].

An intrusion detection system based on GANs was proposed for cyber-physical systems in fog environment [24]. First, data samples about normal patterns were transmitted from the end point layer to the cloud layer. Then, generator and discriminator of GAN structure were trained to synthesize realistic samples and discriminate real patterns from fake ones respectively. By minimizing the mean-squared-error residual loss between real data and the reconstructed one, an encoder was trained within the codec structure where the trained generator was considered as the decoder. The discrimination and reconstruction losses were finally used to detect the intrusion for the evaluated data pattern. To secure the automated vehicle system, GAN was used to establish an anomaly monitoring mechanism by Kim *et al.* [25]. Due to the difficulty of acquiring anomalous data from vision sensors, the generator was trained to learn distribution of the anomalous data. Then large numbers of samples regarding minority anomalies were synthesized. The samples were finally used to train an anomaly detection framework by exploring temporal dependencies between the vehicle motion and vision sensor. To determine whether a transmitter is trustworthy or not, an adversarial learning framework regarding radio frequency signals was proposed by Roy *et al.* [26]. Realistic samples from the trusted transmitters were first collected and used to train the generator of a GAN. Then the counterfeit data was generated from real sample space by impersonating the trusted transmitter. Both the realistic data and the generated one were finally used to train the discriminator where two nodes were deployed in the output layer to determine whether an input was counterfeited.

The applications of GANs in security assurance have been mainly focused on anomaly detection, while trust management with GAN techniques has been rarely addressed. The presented method in [26] has explored the application of GANs in trusted identity detection, whereas, just the original GAN structure has been adopted in this method. Since the absolute classification loss boundary between the real samples and fake ones is hard to determine, false positive detections are easy to happen for the GAN-based anomaly detection mechanisms. To address these issues, the conditional information will first be added to better guide the training of GAN in our proposed method, and then a codec structure will be constructed for trust decision-making. Moreover, to reduce the false positive rate, context information of real samples will be explored for trust prediction.

### III. PROPOSED TRUST MANAGEMENT METHOD WITH GENERATIVE ADVERSARIAL LEARNING

The framework of our generative adversarial learning based trust management (GALTM) method is given in Fig. 1. The procedure is as follows: First, trust evidences that reflect data transmissions are collected in initial network operation phase. Then, fuzzy inferring is performed to estimate the trust values for training dataset preparation. After that, a trust classification model is trained by using the prepared dataset for further trust decision-making. If a node is considered to be trustworthy, then the related trust information will be gathered for further update of the trust classification model. Otherwise, a trust redemption model is established or updated to give false positive nodes the chance to rejoin the network.

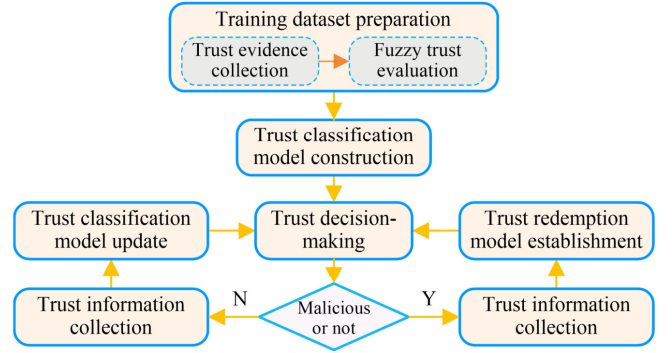

Fig. 1. Framework of the proposed method GALTM.

*A. Trust evidence collection and fuzzy trust evaluation*

To get reliable and real-time communication in IWSNs, trust model should be established for trust decision-making. Then before trust evaluation, trust evidences regarding transmissions need to be collected. Three kinds of transmission attacks exist in the network, including the packet dropping, delaying, and tampering. We consider a cluster-based IWSN where a cluster member first delivers data to the head and then monitors the followed behaviors for evidence collection in each transmission round. The collected transmission evidences are recorded into three binary sequences, where 1 and 0 denote the occurrence and nonoccurrence of the corresponding attacks respectively. As the network operates, more evidences can be recorded.

Due to interferences in industrial environment, node faults or link failures may happen occasionally. Then the collected trust evidences are with some uncertainties as it is hard to determine whether a transmission failure or latency is caused by malicious attacks. To overcome such uncertainty problem and improve the trust evaluation accuracy, interval type-2 fuzzy logic will be introduced for trust evaluation due to its excellent performance in dealing with the fuzziness and randomness [27]. An interval type-2 fuzzy logic system consists of a fuzzy rule base and four modules, which include fuzzifier, fuzzy inferring engine, type reducer, and defuzzifier [6]. To estimate trust values of a node, a sliding window with length $l_{w1}$ is adopted to extract evidences from the sequences regarding this node. Then the trust attribute called packet loss rate is calculated based on packet dropping and tampering evidences, since a tampered packet has to be dropped due to authentication failure. Meanwhile, another trust attribute named transfer delay rate can be computed based on packet delaying evidences. Next, traverse the entire evidence sequences with one-bit step size, some trust attribute pairs can be acquired and finally used to infer the trust values.

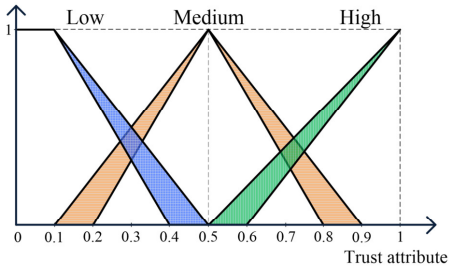

Fig. 2. Membership functions.

TABLE I
FUZZY RULES

| No. ($n$) | Packet loss rate ($x_1$) | Transfer delay rate ($x_2$) | Output trust level ($y_t$) |
|---|---|---|---|
| 1 | Low | Low | 1 |
| 2 | Low | Medium | 0.9 |
| 3 | Low | High | 0.7 |
| 4 | Medium | Low | 0.7 |
| 5 | Medium | Medium | 0.5 |
| 6 | Medium | High | 0.3 |
| 7 | High | Low | 0.3 |
| 8 | High | Medium | 0.1 |
| 9 | High | High | 0 |

The range of each trust attribute can be divided into three overlapping parts, namely [0, 0.5], [0.1, 0.9], and [0.5, 1]. Then three interval type-2 fuzzy sets corresponding to the linguistic variables low, medium, and high can be used to measure the membership degree, as in Fig. 2. To map input trust attribute pair and output trust level of the fuzzy logic system, some Takagi-Sugeno-Kang (TSK) [28] rules are defined using the IF-THEN clauses, as in Table I. Each TSK rule is of the form

$$\tilde{R}^n : \text{IF } x_1 \text{ is } X_1^n \text{ and } x_2 \text{ is } X_2^n, \text{THEN } y_t \text{ is } Y^n, \quad n = 1, \cdots, N. \quad (1)$$

Where $x_1$ and $x_2$ are the input trust attributes, $y_t$ is the output trust level, $X_1^n$ and $X_2^n$ are interval type-2 fuzzy sets regarding the $n$th rule, $Y^n$ is a crisp value regarding the trust level of the $n$th rule, and $N$ is the total number of rules.

Let $x = \langle x_1, x_2 \rangle$ be a general trust attribute pair. To acquire the final trust value via fuzzy inferring, several steps need to be performed accordingly. First, obtain the matching membership degree between the input trust attribute $x_i$ ($i$ = 1, 2) and IF-part of the $n$th ($n$ = 1, 2, ..., $N$) rule, denoted by $[D_L^n(x_i), D_U^n(x_i)]$. Where $D_L$ and $D_U$ are the lower and upper membership degrees, respectively. Next, acquire the firing degree $[D_L^n(x), D_U^n(x)]$, where $D_L^n(x) = T(D_L^n(x_1), D_L^n(x_2))$, $D_U^n(x) = T(D_U^n(x_1), D_U^n(x_2))$, and $T$ is product $t$-norm [27]. Begian–Melek–Mendel (BMM) [28] method is finally adopted to bypass the type reduction and directly calculate the defuzzified output trust value $v_t(x)$:

$$v_t(x) = \alpha \frac{\sum_{n=1}^{N} y_t^n D_L^n(x)}{\sum_{n=1}^{N} D_L^n(x)} + \beta \frac{\sum_{n=1}^{N} y_t^n D_U^n(x)}{\sum_{n=1}^{N} D_U^n(x)}, \quad (2)$$

where $\alpha$ and $\beta$ are adjustable coefficients, $N$ is the total number of rules, $y_t^n$ is the trust level with regard to the $n$th rule, $D_L^n(x)$ and $D_U^n(x)$ are the lower and upper firing membership degrees of $x$ with respect to the $n$th rule.

### B. Trust classification model construction

For the trust decision-making, a classification model usually needs to be created to determine whether a node is trustworthy. However, the lack of experience in some industrial applications makes trust decision-making hard to be performed. In addition, due to harsh and dynamic industrial environments, the inferred trust value of a node may deviate from the reputation that this node deserves. Inspired by the ability of GAN to learn sample distribution, a codec structure can be trained to create the trust classification model. Then trust decision-making can be carried out through determining whether the relevant trust information satisfies sample distribution. Usually, malicious nodes rarely exist in initial network deployment stage. To prepare training dataset, trust values acquired in this stage are used to constitute some trust vectors where each one is a sequence of historical trust values with the length $l_{w1}$. These trust vectors are regarded as the real samples and further used to train the codec structure.

In contrast to the original GAN, the conditional generative adversarial network (CGAN) is more controllable since some conditional information is introduced to better guide the sample distribution learning [14], [16]. Then a GAN pair that includes an original GAN and a CGAN is adopted to construct the codec structure, as shown in Fig. 3. Generators of the original GAN and CGAN respectively act as the encoder and decoder, which are the two key components of the codec structure. The encoder learns the latent distribution, and then generates latent data for a trust vector sample. The decoder learns a mapping between the latent space and real sample space conditioned on the sample features. As for the discriminators, one is trained to distinguish real data from the fake one regarding the latent space, while another is trained to decide whether a conditioned data satisfies real sample distribution. The original GAN is optimized by the adversarial $min$-$max$ game between generator and discriminator [15], and the objective function is of the form

$$\min_G \max_D V(G, D) = E_{p_{sample}(x)}[\log D(x)] + E_{p_{latent}(z)}[\log(1 - D(G(z)))], \quad (3)$$

where $x$ is the real sample, $p_{sample}$ is the sample distribution, $D(x)$ denotes the output score for $x$ marked by the discriminator, $z$ is the latent data, $p_{latent}$ stands for the latent distribution, $G(z)$ is the generated data for $z$, and $E$ denotes the expectation.

As for CGAN, conditional information $c$ is introduced into the adversarial game [16]. Then the optimization objective can be formulated by

$$\min_G \max_D V(G, D) = E_{p_{sample}(x)}[\log D(x|c)] + E_{p_{latent}(z)}[\log(1 - D(G(z|c)|c))]. \quad (4)$$

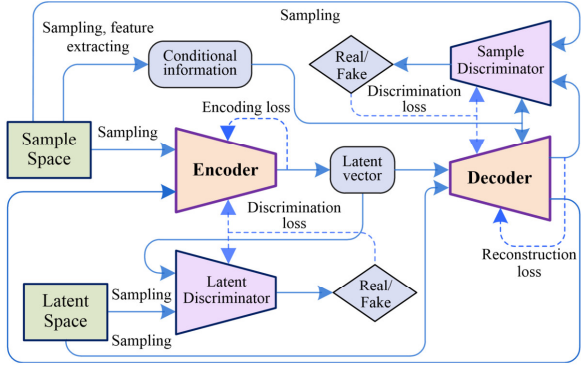

Fig. 3. The codec structure.

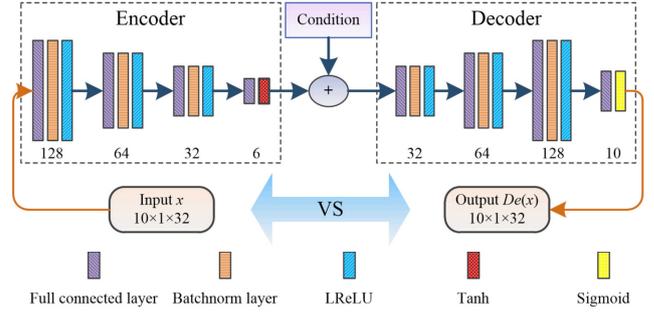

Fig. 4. Detailed codec architecture.

The detailed codec architecture is given in Fig. 4. Four fully connected layers are included for both the encoder and decoder. Since a trust value should range from 0 to 1, *Sigmoid* is adopted as the activation in the last layer of the decoder. Moreover, the changes of historical trust values are expected to be reflected in the latent space. Then in the last layer of the encoder, *tanh* is used as the activation since the difference between two trust values should range from -1 to 1. To train the codec structure, batch normalization method is adopted to improve the stability and convergence speed. In addition, trust vector size is set to be 10×1, and batch size is 32. In each training epoch, a batch of trust vector samples is first selected and fed into the encoder. The conditional information, which reflects the changes of trust values for some randomly selected samples, is appended to the output of the encoder and then fed into the decoder to guide data generating. After that, a batch of randomly selected latent data with conditional information is also fed into the decoder for data synthesizing. The synthesized data is then encoded by the encoder to get the corresponding data that satisfies the latent distribution. Finally, a back propagation is performed for the corresponding losses, and then the parameters of the codec structure are updated using the *Adam* optimizer. The loss of the encoder, which is expressed as $L_{En}$, consists of two parts: One is the discrimination loss that can be defined as the *Least squares* between the expected output and the actual output of the latent discriminator. Another is the encoding loss, which is defined as the mean absolute difference between the latent data and the corresponding encoded data. Then $L_{En}$ can be formulated by

$$L_{En} = \frac{1}{2}E_{p_{sample}(x)}\left[\left(D_L(En(x))-1\right)^2\right] \\ + \frac{1}{2}E_{p_{latent}(z)}\left[\left|z-En(De(z|c))\right|\right], \quad (5)$$

where $x$ is a real sample data, $z$ is a latent data, $c$ is a conditional information, $En(x)$ is the encoded data for $x$, $D_L(\cdot)$ is the score marked by the latent discriminator, $De(\cdot)$ is the data generated by the decoder.

In addition to the loss of *Least squares* between the expected output and the actual output from the sample discriminator, the decoder has a reconstruction loss, which is defined as the mean absolute difference between the input and output of the codec. Then the loss of decoder, expressed as $L_{De}$, can be expressed by

$$L_{De} = \frac{1}{2}E_{p_{latent}(z)}\left[\left(D_S(De(z|c)|c)-1\right)^2\right] \\ + \frac{1}{2}E_{p_{sample}(x)}\left[\left|x-De(En(x)|c)\right|\right], \quad (6)$$

where $D_S(\cdot)$ is the score marked by the sample discriminator.

After the codec is well trained, the current trust vector of a node can be first built and then fed into the codec. Following that, by computing the reconstruction loss of the codec, whether this node is trustworthy or not can be determined.

### C. Trust redemption model establishment

Due to various interferences in industrial field, false positive error may occur for the malicious nodes detection. Then a trust redemption model should be established to avoid false positive nodes being permanently isolated from the network. Owing to the capability of learning sample distribution, GAN has the potential to restore the masked samples. Therefore, GAN can be trained to predict attacks from a node that is considered as the false positive, and then whether the trust redemption is available for this node can be determined. Based on this view, we present a GAN-based trust redemption model, and details will be discussed below.

Unlike the trust classification model, trust redemption model is initially established in the steady network operation phase. For preparing model training dataset, the binary trust evidence sequences of the nodes that are regarded as the malicious need to be collected in this phase. The sequence size is set to be $l_{w1}$ to allow only some latest transmission evidence records be left. The binary evidence sequences of a node are first fused into a single one. If a bit value is 1 in any original evidence sequence, then the value of the corresponding bit in the fused sequence is 1. Next, a sliding window with size $l_{w2}$ is introduced to extract data from the fused evidence sequence. Based on the data from one extraction, an attack probability can be calculated. Traverse the entire evidence sequence with one-bit step size, multiple attack probabilities can be obtained to form a vector. As the network operates, if the number of prepared probability vectors regarding the malicious reaches ten batches, then the dataset is ready for training the trust redemption model.

The presented GAN-based trust redemption model, which consists of a generator and discriminator, is given in Fig. 5. The generator can be further divided into an encoder and a decoder, which are trained to extract the feature of the masked sample and generate data according to that feature, respectively. The discriminator is trained to distinguish the real sample from the restored one. In each training epoch, a batch of real samples is selected from the prepared dataset, and the last digit of each real sample is covered by a random mask. Then the masked samples are fed into the generator. Next, to acquire the restored samples, the last digits of the generated samples are used to replace the masks of the masked samples. Finally, the restored samples and another batch of real samples are assigned the scores by the discriminator, and the losses are fed back to update both the generator and discriminator.

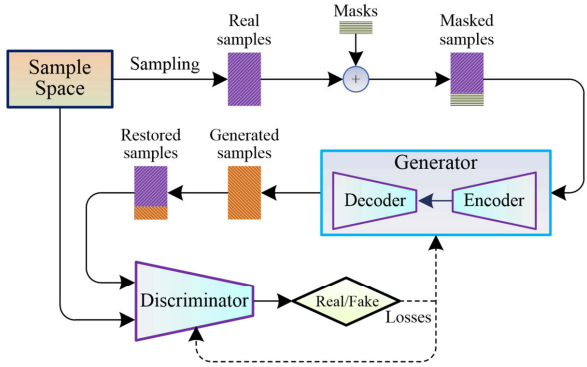

Fig. 5. GAN-based trust redemption model.

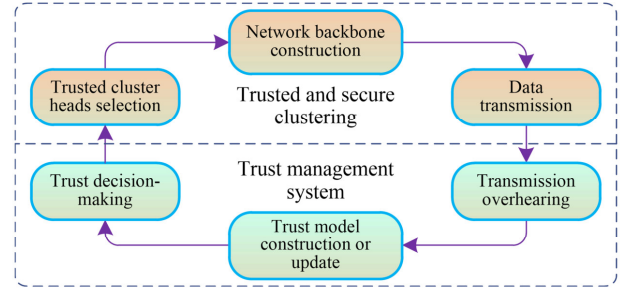

Fig. 6. The proposed trusted and secure clustering model.

The well trained trust redemption model can be adopted to predict the future attack probability, which is considered as the mask digit of a real sample. Then the probability of cooperation for a false positive node can be predicted, and the result is used to determine whether this node can be remedied in future.

### D. Trust decision-making and trust model update

To decide whether a node can be trusted, the current trust vector of this node should first be obtained. Then the trust vector is fed into the well trained trust classification model. Next, the reconstruction loss of the trust vector is measured by the mean absolute difference between the input and output. If the reconstruction loss is lower than a threshold $Tr_1$, it indicates that the output does not deviate from the input. Therefore, the input trust vector can be considered to satisfy the real sample distribution while the corresponding node is considered to be trustworthy. Otherwise, the node is considered as the malicious. To avoid a normal node being permanently isolated from the network due to error trust decision-making, trust redemption mechanism should be activated if necessary. A node that is detected to be malicious can be treated as a false positive node if both the following two conditions are satisfied: First, the reconstruction loss for the trust vector of this node is higher than the threshold $Tr_1$ but lower than another bigger threshold $Tr_2$. Second, there are no fewer than three trusted neighbors of the decision node can recommend this node as a trusted one. To provide an opportunity of rejoining network for the node that is considered to be false positive, future cooperation probability of this node first needs to be predicted using the trained trust redemption model. Then this node can finally be trusted by others with the cooperation probability and would not be directly isolated from the network.

The above thresholds $Tr_1$ and $Tr_2$ for trust decision-making are determined dynamically. First, compute the reconstruction losses for training samples according to the well trained trust classification model. Then record all the losses into a sequence, and the maximum loss is regarded as the threshold $Tr_2$. Next, remove the 10% maximum losses in the sequence, and finally the maximum of the remaining is considered as the threshold $Tr_1$. Once the trust classification model is updated, the two trust thresholds $Tr_1$ and $Tr_2$ also need to be recomputed.

Due to dynamic industrial environment, the initially trained trust model may be ineffective as the network operates. Then it is crucial to update the trust model appropriately in the future network operation. For updating the trust classification model, a dataset that consists of some latest trust vectors needs to be built. If a testing node is detected to be trustworthy while the reconstruction loss for its trust vector is lower than 0.6 times the threshold $Tr_1$, then this trust vector can be assigned to that dataset. Once the dataset size reaches five batches, the update of the trust classification model can be performed via several times of retraining. Similarly, if a testing node is detected as the malicious, the corresponding attack probability vector is stored into another dataset. Once the dataset size reaches five batches, the update of the trust redemption model can be activated.

## IV. THE PROPOSED TRUSTED AND SECURE CLUSTERING

Clustering is usually adopted in sensor networks for network backbone construction due to its excellent performance in energy efficiency and scalability. For secure data delivery, we present a trusted and secure clustering protocol where the trust management method GALTM is used by sensor nodes for trust decision-making and reliable cluster heads selection.

The proposed clustering model is given in Fig. 6. It can be divided into two parts: One is the trust management system, and another is the clustering module. The trust management system can guide the construction of reliable network backbone, while the clustering can provide the newest trust information for trust model update. Due to high computational complexity of GAN training, several trust agencies are used for initial trust models construction. After that, each sensor node decides being the cluster head based on a threshold $Tr$ if it is eligible:

$$Tr = \frac{p_c}{1 - p_c \left( r \bmod \left( 1/p_c \right) \right)}, \qquad (7)$$

where $p_c$ is the probability of being the head, $r$ is the current round number. If a sensor node has not been a head in the latest $1/p_c$ rounds, it is currently eligible to be a head.

Then, based on the trust model shared by the nearest agency, a non-cluster head performs trust decision-making to select the nearest trusted neighbor head to join the cluster. A node that does not have a trusted neighbor head has to be the head if it is eligible. Once the cluster formation phase ends, a cluster-based network backbone is constructed. Next, each member node first collects and transmits data to the head within an individual time slot, and then overhears the followed transmission to confirm whether its data is timely forwarded without tampering. Finally, each node updates its own trust model by retraining the GANs using some latest trust information. The scale of training dataset for trust model update should be small enough to limit the computational complexity.

## V. Algorithm Performance Analysis

In this section, we will theoretically analyze the performance of our trust management method in terms of the computational complexity, communication cost, and security assurance.

### A. Computational complexity

For our trust management method, the main computational complexity comes from model training. It has been proved that a neural network with depth $\sigma$ can be learned in $poly(D^{2^\sigma})$ time [26], here $D$ is the dimension of input and $poly(\cdot)$ is a constant time that depends on system configuration. In our method, the trust classification and trust redemption models are constructed using GANs where both the generator and discriminator have 4 layers. As the trust classification model has 2 pairs of generator and discriminator while an additional discriminator exists in the trust redemption model, the total layers of the trust model in our method is $4 \times 9 = 36$. The dimension of dataset used for initial trust model training is about 10K, and that for each time of trust model update is 160, Then the computational complexities can be estimated by $poly(10e3^{2^{36}})$ and $poly(160^{2^{36}})$ for initial trust model training and each time of model update, respectively.

Since GAN is a kind of deep learning technique, it has higher complexity than other traditional machine learning techniques, such as SVM and C4.5 decision tree that are respectively used to construct the trust models in [9] and [11]. The computational complexity of SVM is usually between $poly(N_s^3 + DN_s^2 + DN_s d_L)$ and $poly(D^2 d_L)$, where $N_s$ is the number of support vectors, $D$ is the training dataset dimension, and $d_L$ is the size of each sample. The computational complexity of C4.5 is $poly(DN_f D_p)$, where $N_f$ is the number of features and $D_p$ is the depth of decision tree.

### B. Communication cost

In our method, the initial trust models are trained by several trust agencies. Then each sensor node needs to communicate with the nearest trust agency once to acquire the trained model parameters. A sensor node may have additional communication cost in each data transmission round for selecting the trusted cluster head to join cluster. If a node thinks the cluster head to be selected may be a false positive node, it communicates with several trusted neighbors once for further decision. However, this case only happens occasionally and the frequency depends on the stability of communication environment. Other methods like [19] and [20] have higher communication cost as the trust recommendation is necessary for every trust decision-making.

### C. Security analysis

Considering that the result of a single trust evaluation may be inaccurate, the trust vector that includes the number $l_{w1}$ of trust values is used for trust decision-making in the proposed method. Then if a node $i$ needs to determine whether another neighbor $j$ is trustworthy, it should have directly interacted with $j$ at least $l_{w1}$ times for evidences collection and trust values estimation. Node $i$ can directly interact with $j$ only when they are in the same cluster where $j$ acts as the cluster head, then it needs to wait for an average of $N_r$ data transmission rounds for the first time of trust decision-making. Let $S$ be the network size, $R$ be the neighborhood radius, and $n$ be the total number of nodes in the network. Since $i$ selects the cluster head from its neighbor nodes, it chooses $j$ as the head with the probability $S/(\pi R^2 n p_c)$ if $j$ decides to be the head, where $p_c$ is the probability of being a head. Then $N_r$ can be estimated by

$$N_r = \frac{l_{sw1}}{p_c} \frac{\pi R^2 n p_c}{S} = \frac{l_{sw1} \pi R^2 n}{S}. \qquad (8)$$

For a longer trust vector, trust decision-making has to be performed later while the accuracy can be improved. Many other standard trust models such as [9] and [18] can be used to detect malicious nodes earlier as a single trust value is directly used for trust decision-making, whereas, the detection accuracy highly depends on the current trust value. Moreover, owing to the adversarial learning ability of GAN, our trust management method can rapidly adapt to dynamic industrial environments since only a small amount of data needs to be collected for trust model update. Other trust models such as [11] and [22] cannot achieve rapid updates since more trust information is necessary.

## VI. Experiment Evaluation and Analysis

### A. Experimental Setup

In our experiment, totally 100 sensor nodes are deployed into an industrial field with size $100 \times 100$ m$^2$. Three kinds of the malicious called the normal, advanced, and super nodes exist in the network, and they launch both dropping (tampering) and delayed forwarding attacks with probabilities $P_{at}$, $2P_{at}$, and $3P_{at}$, respectively. The percentages of these nodes in total number of malicious ones are 30, 40, and 30, respectively. To reflect the impact of node faults and link failures on data transmission, a Markov chain is used to indicate the communication status that is either bad or good with the probability $p_0$ [13]. It is assumed that a data packet cannot be successfully captured for a bad status. The dataset for initial trust model training is gathered in the initial network deployment stage, since malicious nodes rarely exist in this stage.

Cloud model and fuzzy logic theory have been used for fuzzy trust evaluation in TECC and TEUC, respectively. Especially, cloud model is very adaptive for uncertainty description since the trust uncertainty degree can be measured by cloud thickness. TEUC is an excellent trust management method with machine learning techniques. C4.5 decision tree is adopted to learn trust patterns and construct the trust model, while the issue of trust model update in dynamic environment is addressed. As both the trust uncertainty problem and the intelligent trust model construction and update issue have been studied in the proposed method GALTM, we verify the performance of GALTM by comparing it with that of TECC and TEUC. For development of the followed experiments, dropping, delaying, and tampering events are also captured as the trust evidences in TECC and TEUC. Failure tolerance factor is set to be 0.2 while time sensitive factor is 0.6 in TECC. The thresholds for evidence fuzzification are 0.9 and 0.7 in TEUC. Parameter settings with regard to energy consumption are the same with our previous work [13], and others are given in Table II.

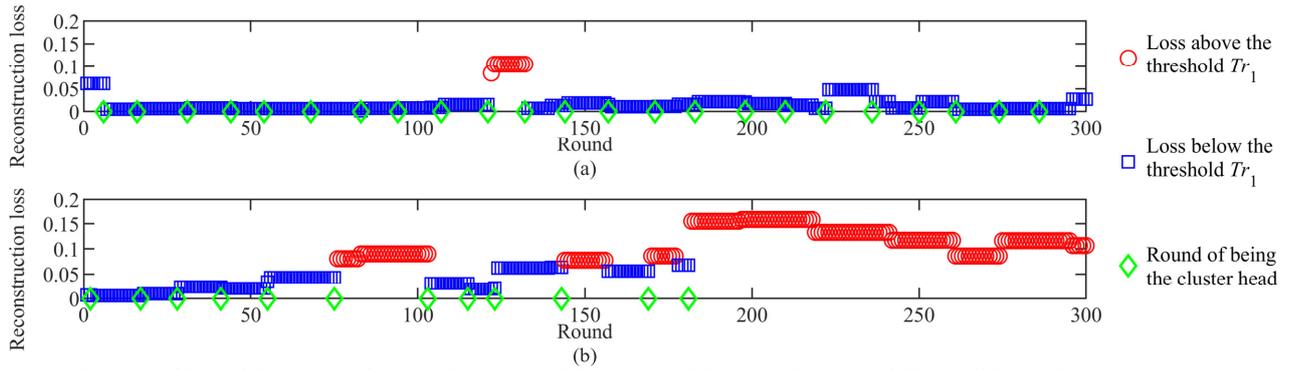

Fig. 7. Some instances of the real-time adversarial scenario. a) For a neighbor non-malicious node; b) For a neighbor malicious node.

TABLE II
PARAMETER SETTINGS

| Parameter | value |
| --- | --- |
| Packet size (bits) | 3000 |
| Control packet size (bits) | 300 |
| Initial energy $E_0$ (J) | 1.3 |
| Probability $p_c$ of being cluster heads | 0.07 |
| Neighborhood radius $R$ (m) | 25 |
| Probability $P_{at}$ of malicious attacks | 0.1 |
| Probability $p_0$ of the bad communication status | 0.1 |
| Length $l_{w1}$ of sliding window (bits) | 10 |
| Length $l_{w2}$ of sliding window (bits) | 4 |
| Adjustable coefficients $\alpha$ and $\beta$ for defuzzification | 0.5 |

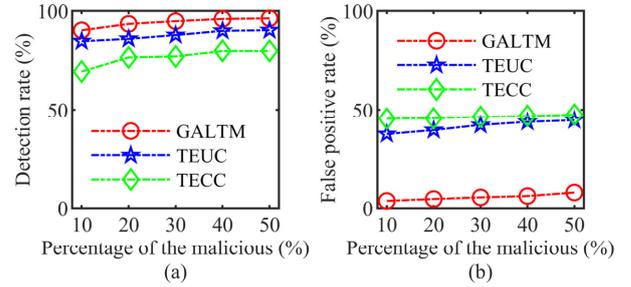

Fig. 8. Comparisons of the detection performance. a) Detection rate; b) False positive rate.

### B. Experiment Results

To show the changes in trust relationship and the final trust decision, some instances of the real-time adversarial scenario are given in Fig. 7. The results of trust decision-making using the proposed trust classification model during totally 300 data transmission rounds are illustrated, while a) and b) are for the neighbor non-malicious and malicious nodes of the decision node, respectively. A green diamond in the figure denotes the node has been selected as the cluster head in the corresponding round. A blue square represents the reconstruction loss of our trust classification model for a trust vector of the node is below the threshold $Tr_1$ while this node is considered to be trustworthy, and a red circle means the reconstruction loss is above $Tr_1$ in the corresponding round. From a) we can see that a non-malicious node can always be selected as the cluster head at intervals, which indicates that the decision node keeps trust in this node. Although the reconstruction loss regarding the non-malicious node has ever exceeded the threshold $Tr_1$ in some rounds, this node has gained the opportunity for trust redemption in the subsequent rounds since the loss is always smaller than another threshold $Tr_2$. On the contrary, a malicious node can finally be identified and then isolated from being cluster head according to b). Although the malicious node has ever obtained several opportunities for trust redemption and then acted as the cluster head, the opportunity for trust redemption is finally no longer available for this node. This is because trust redemption is possible only if both the following two conditions are satisfied: One is that the related reconstruction loss must be below the threshold $Tr_2$. Another is that there are no fewer than 3 trusted neighbors of the decision node can provide recommendations.

The comparisons of detection rate and false positive rate are given in Fig. 8. Here detection rate is defined as the ratio of correct detection times to total detection times for the malicious nodes, and false positive rate is defined as the ratio of error detection times to total detection times for the non-malicious nodes. The result denotes that the proposed method GALTM performs better in terms of detection and false positive rates than TECC and TEUC. Especially, the detection rate is up to 96% while the false positive rate is lower than 8% even if the percentage of the malicious nodes is 50. This is because GAN can better learn the characteristics of samples than traditional machine learning methods. Moreover, owing to the presented trust redemption model, the improvement of performance in false positive rate is more obvious than that in detection rate for our method GALTM. This figure also shows that for the three methods, both detection and false positive rates increase slowly when the percentage of the malicious increases. This is because that the total number of malicious nodes increases while that of non-malicious ones decreases if percentage of malicious nodes increases. Then more chances are available for the detection of malicious nodes, whereas, error detection times decrease more slowly than the total detection times for non-malicious nodes since the former is much smaller than the latter.

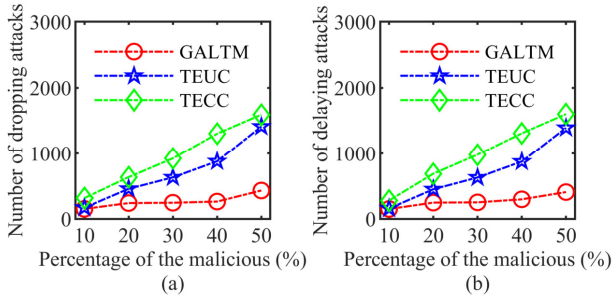

Fig. 9. Comparisons of total number of attacks. a) Packet dropping attack; b) Packet delaying attack.

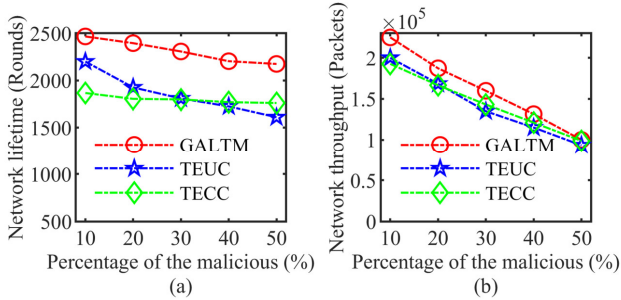

Fig. 10. Comparisons of the network performance. a) Network lifetime; b) Network throughput.

The comparisons of number of attacks are given in Fig. 9. It shows that the number of attacks increases if the percentage of the malicious increases. When the percentage of the malicious is 10%, GALTM does not outperform TECC and TEUC in preventing against the attacks. This is because a sensor node in the network using GALTM needs some time to get the trust vector before the first time of trust decision-making, meanwhile, GALTM has a better resilience since a trust redemption model is used to give the nodes considered to be false positive the chance to rejoin the network. Owing to the adversarial learning mechanism, GALTM has better performance in preventing against attacks than TECC and TEUC for a percentage of the malicious bigger than 10. Especially, GALTM reduces attacks by more than 70% if the percentage of the malicious is 50.

The comparisons of lifetime and throughput of the network are given in Fig. 10. We define network lifetime as total alive rounds of the non-malicious node that first exhausts the energy. The network throughput is defined to be the total number of packets that are collected by non-malicious sensor nodes and finally have been successfully delivered to the destination. This figure shows that the network with the higher percentage of malicious nodes has shorter lifetime and smaller throughput as more malicious nodes are detected and fewer non-malicious nodes are left to burden heavier delivery tasks. Owing to our trust redemption model, the network using GALTM has longer lifetime than those using TECC and TEUC. Especially, the network lifetime is increased by more than 19% if the percentage of malicious nodes is 50, since false positive nodes still have chances to join the network to provide services if our method GALTM is used. Then the resource utilization of non-malicious nodes can be improved so that GALTM can perform better in terms of network throughput.

## VII. CONCLUSION

This paper has presented a GAN-based trust management framework, which includes the courses of trust evaluation, trust classification, trust redemption, and trust model update. The proposed method can effectively alleviate the trust uncertainty problem caused by node faults or link failures in industrial environment. Especially, it can efficiently detect the malicious due to the potential of GANs to learn the distribution of samples. Moreover, the proposed method has a good resilience in trust management and the trust model is timely updated to adapt to the industrial environment. Simulation results have verified its excellent performance in securing IWSNs.

In future work, we will further study the GAN-based trust model by using the supervised or semi-supervised methods to accelerate the training speed. Meanwhile, applications of other emerging machine leaning techniques like the Q-Learning and Federated Learning in trust management will also be explored. Moreover, to improve trust management efficiency, blockchain will be considered for reliable trust recommendation.

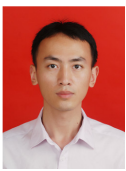

**Liu Yang** received his B.S. degree in Electronic Information Science and Technology from Qingdao University of Technology, Shandong, China, in 2010, and Ph.D. degree in Communication and Information Systems at the College of Communication Engineering, Chongqing University, Chongqing, China, in 2016. He is now a lecturer in Chongqing University of Posts and Telecommunications. His research interests include Internet of Things, data analysis, and artificial intelligence.

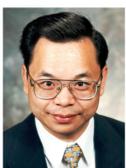

**Simon X. Yang** (S'97–M'99–SM'08) received the B.Sc. degree in engineering physics from Beijing University, Beijing, China, in 1987, the first of two M.Sc. degrees in biophysics from the Chinese Academy of Sciences, Beijing, China, in 1990, the second M.Sc. degree in electrical engineering from the University of Houston, Houston, TX, in 1996, and the Ph.D. degree in electrical and computer engineering from the University of Alberta, Edmonton, AB, Canada, in 1999.

Dr. Yang is currently a Professor and the Head of the Advanced Robotics and Intelligent Systems Laboratory at the University of Guelph, Guelph, ON, Canada. His research interests include robotics, intelligent systems, sensors and multi-sensor fusion, wireless sensor networks, control systems, machine learning, fuzzy systems, and computational neuroscience.

Prof. Yang has been very active in professional activities. He serves as the Editor-in-Chief of International Journal of Robotics and Automation, and an Associate Editor of IEEE Transactions on Cybernetics, IEEE Transactions on Artificial Intelligence, and several other journals. He has involved in the organization of many international conferences.

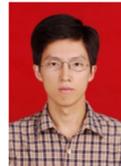

**Yun Li** is currently a professor with the School of Software Engineering, Chongqing University of Posts and Telecommunications, Chongqing, China. He received his Ph.D. degree in communication engineering from the University of Electronic Science and Technology of China. His research interests include mobile cloud/edge computing, cooperative/relay communications, green wireless communications, wireless ad hoc networks, sensor networks, and virtual wireless networks. He is the Executive Associate Editor of Elsevier/CQUPT Digital Communications and Networks (DCN). He is on the editorial boards of IEEE Access and Security and Communication Networks.

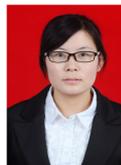

**Yinzhi Lu** received her M.S. degree in Communication and Information Systems from Chongqing University, Chongqing, China, in 2014. She is currently pursuing the Ph. D. degree in Information and Communication Engineering with the School of Communication and Information Engineering, Chongqing University of Posts and Telecommunications. She was a teaching assistant with the School of Electronic Information Engineering, Yangtze Normal University from 2014 to 2019. Her current research interests include Internet of Things, time sensitive network, and artificial intelligence.

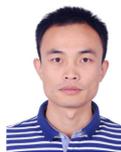

**Tan Guo** (S'17–M'22) received his M.S. degree in Signal and Information Processing from Chongqing University, Chongqing, China, in 2014, and Ph.D. degree in Communication and Information Systems from Chongqing University, Chongqing, China, in 2017. He is now a lecturer in Chongqing University of Posts and Telecommunications. His research interests include Internet of Things, biometrics, pattern recognition, and machine learning.